\documentclass[preprint,11pt]{elsarticle}

\usepackage{amssymb}
\usepackage{graphicx}
\usepackage{amsfonts}
\usepackage{amsmath}
\usepackage{amssymb}
\usepackage{bbold} 
\usepackage[margin=.7in]{geometry}
\usepackage{booktabs}
\usepackage{float}
\usepackage{mathrsfs}
\usepackage{hyperref}

\journal{Nuclear Physics B}

\usepackage{lscape}
\usepackage{xcolor}
\usepackage{booktabs}
\usepackage{float}
\usepackage{mathrsfs}
\usepackage{hyperref}
\usepackage{multirow}

\newcommand{\UPMNS}{U_{\mathrm{PMNS}}}

\newcommand{\ord}{\mathcal{O}}
\newcommand{\mbeta}{m_{\beta}}
\newcommand{\mbetabeta}{m_{\beta \beta}}
\newcommand{\ths}{\theta_{12}}
\newcommand{\thr}{\theta_{13}}
\newcommand{\tha}{\theta_{23}}
\newcommand{\chisquare}{\chi^2}

\newcommand{\mbf}{\mathbf}

\newcommand{\de}{\mathrm{d}}
\newcommand{\Ke}{\mathcal{K}}

\newcommand{\PLANCKCON}{PLANCK TT + lowP + Lensing}

\newcommand{\PLANCKAGG}{PLANCK TT + lowP + Lensing + Ext} 

\begin{document}

\begin{frontmatter}



\title{Probability Densities of the effective neutrino masses $m_{\beta }$ and $m_{\beta \beta}$ }


\author[]{Andrea Di Iura}
\ead{diiura@fis.uniroma3.it}
\author[]{Davide Meloni}
\ead{meloni@fis.uniroma3.it}

\address{Dipartimento di Matematica e Fisica, Universit\`a  di Roma Tre;\\
INFN, Sezione di Roma Tre,\\
Via della Vasca Navale 84, 00146 Rome, Italy}

\begin{abstract}

We compute the probability densities of the effective neutrino masses $m_{\beta }$ and $m_{\beta \beta}$ 
using the Kernel Density Estimate (KDE) approach applied to a distribution of points in the 
$(m_{\min}, \mbetabeta)$ 
and $(\mbeta, \mbetabeta)$ planes, 
obtained using the available Probability Distribution Functions (PDFs) of the neutrino mixing angles and 
mass differences,  with the additional constraints coming from cosmological data on the sum of the neutrino masses.
We show that the reconstructed probability densities strongly depend on the assumed set of cosmological data: 
for $\sum_j m_j \leq 0.68$ eV at $95\% \ \mathrm{CL}$ a sensitive portion of the allowed values are already excluded by null results  
of experiments searching for  $m_{\beta \beta}$ and $m_{\beta }$,
whereas in the case $\sum_j m_j \leq 0.23$ eV at $95\% \ \mathrm{CL}$  the bulk of the probability densities are below the current bounds.
\end{abstract}


\end{frontmatter}

\section{Introduction}
Although the physics of neutrino oscillation is entering a precision era, with all mixing angles and absolute values of the mass differences 
measured at the level of some percent, there are still questions related to the nature of neutrinos that need to be answered.
Among these, we are interested in
whether neutrinos are Majorana or Dirac particles and in the absolute value of their masses. As it is well known, experiments on neutrinoless double beta decays ($0\nu\beta\beta$) 
consider the possibility that the reaction 
\begin{eqnarray}
( A , Z ) \longrightarrow (A , Z + 2) + e^-+ e^-\, \label{process} 
\end{eqnarray}
really occurs; in the case of positive signal, we could conclude that the total lepton number is violated by two units, although the process 
behind the conversion of two down quarks into two up quarks would not be uniquely determined
 \cite{Furry:1939qr,Duerr:2011zd, Vergados:2016hso}.
The $0\nu\beta\beta$-decay amplitude has the form
$ \mathcal{A}(0\nu\beta\beta) = \mbetabeta \mathcal{M}(A, Z)$,
where $\mathcal{M}(A, Z)$ is the nuclear matrix element of the decay in Eq.~\eqref{process} that does not depend on the neutrino masses and mixing parameters, and  $m_{\beta \beta}$ is the effective mass which, in the case of three lepton families, is given by
\begin{align}
\label{mbetabeta}
 m_{\beta \beta} \equiv \bigg|\sum_j m_j U_{ej}^2 \bigg|= \bigg|\cos^2\theta_{13} \Big(m_1 \cos^2\theta_{12} +m_2 \sin^2\theta_{12}e^{i \alpha}  \Big)+ m_3\sin^2\theta_{13} e^{i \beta}\bigg|.
\end{align}
In the previous relation, $ U_{ej}$ are the elements of the Pontecorvo-Maki-Nakagawa-Sakata (PMNS)
mixing matrix $\UPMNS$ that encodes the leptonic mixing angles $\theta_{ij}$, whereas the phases $\alpha$ and $\beta$ are the so-called 
Majorana phases (one of which eventually absorbs the $CP$ violating phase $\delta$).
As it is usually done, two of the three neutrino masses $m_j$ in Eq.~\eqref{mbetabeta} can be expressed in terms of the lightest one $m_{\min}$  in a way 
that dependents on the supposed neutrino mass hierarchy; for Normal Ordering (NO) we have:
\begin{equation}
m_1 = m_{\min} \qquad m_2 = \sqrt{m_{\min}^2 + \Delta m^2_{21}} \qquad m_3=\sqrt{m_{\min}^2 + \Delta m^2_{31}}  \,,
\end{equation}
whereas for the Inverted Ordering (IO) we set:
\begin{equation}
m_1 =\sqrt{m_{\min}^2 - \Delta m^2_{21}-\Delta m^2_{32}}  \qquad m_2=\sqrt{m_{\min}^2 -  \Delta m^2_{32}} \qquad  m_3 = m_{\min} \,,
\end{equation}
so $m_{\beta \beta}$ effectively depends on the seven independent parameters $\theta_{12}, \theta_{13}, \Delta m^2_{21}, \Delta m^2_{31},\alpha,\beta$ and $m_{\min}$.

The study of the electron spectrum near the end point in the nuclear reaction $^3\mathrm{H} \to  {^3\mathrm{He}} + e + \overline{\nu}_e$
allows, in presence of neutrino mixing, to get information on the other effective mass largely studied in the literature, $m_\beta$, defined by
\begin{align}
\label{mbeta_definition}
  m_{\beta } \equiv \sqrt{\sum_j m_j^2 |U_{ej}|^2 }= \sqrt{\cos^2\theta_{13} \Big(m_1^2 \cos^2\theta_{12} +m_2^2 \sin^2\theta_{12} \Big)+ m_3^2\sin^2\theta_{13} }.
\end{align}
Since absolute values of the PMNS matrix are taken, complex phases play no role and  $m_\beta$ only depends on three independent observables and it is 
somehow correlated, although not in a simple form, with $m_{\beta \beta}$. It is customary to present such a correlation varying all mixing parameters inside 
their 1, 2 or 3$\sigma$ range ($[0,2\pi]$ for the Majorana phases in any case) and computing the maximum and minimum allowed value. While this procedure certainly 
gives insights on the possible outcomes of an experimental search, no information whatsoever can be drawn on the probability distribution
of the observable itself. So, inspired by the work of \cite{Benato:2015via} and \cite{Gerbino:2016ehw}, we computed the  distributions of  $m_{\beta }$ and 
$m_{\beta \beta}$ and the Credible Regions (CR) as obtained using the available PDFs of $\theta_{12}, \theta_{13}, \Delta m^2_{21}$ and $\Delta m^2_{31}$,
with the additional constraints coming 
from cosmological data on the sum of the neutrino masses (see also Refs. \cite{Feruglio:2002af, Dell'Oro:2015tia, Gerbino:2015ixa} and \cite{Ge:2016tfx}). 
However, unlike the procedure 
followed in \cite{Benato:2015via} and \cite{Gerbino:2016ehw}, we use the  KDE approach to compute PDFs of the observables in the $2D$ planes 
$(m_{\min}, \mbetabeta)$ 
and $(\mbeta, \mbetabeta)$.  The use of such a procedure also allows us to save computation time which could become a critical aspect in this sort of simulations.
%

A short summary of the numerical procedure and the data set we used to get the PDFs from the available data 
is done in Sect. \ref{sec:numerical_procedure} whereas a short description of the KDE method and the obtained results 
is done in Sect. \ref{sec:results}. Finally in Sect. \ref{discussion} we compare the PDFs derived  from several 
choices of the Kernel function.

\section{Numerical procedure and datasets}
\label{sec:numerical_procedure}


The construction of the PDFs for $m_{\beta}$ and $m_{\beta \beta}$ passes through the extraction 
of the observables  $p = \{\sin^2\theta_{12},\sin^2\theta_{13}, \Delta m^2_{21}, \Delta m^2_{3\ell} \}$ 
(with $\ell = 1$ for NO and  $\ell = 2$ for IO) from which they depend; the sampling is based on 
the knowledge of the likelihoods $\mathcal{L}(p)$ which in turn are functions
to the single $\Delta \chi^2(p)$:
\begin{align}
\label{like_from_deltachi}
 \mathcal{L}(p) \propto \exp\left( -\frac{\Delta \chi^2(p)}{2}\right) \,.
\end{align}
For the observables $p$ (which are only midly correlated), this information is 
available online at the address
\href{http://www.nu-fit.org}{\texttt{http://www.nu-fit.org}}, where the $\Delta \chi^2$  
for the  November 2016 data, based on the procedure discussed in Ref. \cite{Esteban:2016qun}, are given. 
Notice that a Bayesian analysis on the 2014 data set is available in Ref. \cite{Bergstrom:2015rba}; the authors 
found that the results generally agree (at the level of one standard deviation) with those of the frequentest method, with some differences involving the atmospheric angle $\theta_{23}$ and the Dirac $CP$ violating phase. However, $\theta_{23}$ does not enter into the expressions of the effective masses and the information on the $CP$ phase is hidden by the presence of the Majorana phases. We then decided to use the most recent data set.
%
For the sake of completeness, we report in  Tab. \ref{tab:parameter_table} the central values and $3\sigma$ errors for all the observables relevant in neutrino oscillation for both orderings; similar values 
are also obtained in Ref. \cite{Capozzi:2016rtj}.
In addition to the oscillation data, our estimate of the PDFs also takes into account the cosmological constraints on the sum of the neutrino masses 
$\sum_j m_j$ coming from the Planck experiment \cite{Ade:2015xua}.

\begin{table}[h!]
\begin{center}
\begin{tabular}{c  c  c  c c}
\toprule
\toprule
& \multicolumn{2}{c}{Normal Ordering} & \multicolumn{2}{c}{Inverted Ordering}\\
\midrule
Parameter & Best Fit & 3$\sigma$ Range & Best Fit & 3$\sigma$ Range\\ 
\midrule
$\sin^2\ths/10^{-1}$ & 3.06$^{+0.12}_{-0.12}$ 	 & 2.71 $\div$ 3.45 	& 3.06$^{+0.13}_{-0.12}$ 	& 2.71 $\div$ 3.45 \\
$\sin^2\thr/10^{-2}$ & 2.166$^{+0.075}_{-0.075}$ & 1.934 $\div$ 2.392 	& 2.179$^{+0.076}_{-0.076}$ 	& 1.953 $\div$ 2.408 \\
$\sin^2\tha/10^{-1}$ & 4.41$^{+0.27}_{-0.21}$ 	 & 3.85 $\div$ 6.35 	& 5.87$^{+0.20}_{-0.24}$ 	& 3.93 $\div$ 6.40\\
 $\delta$ 	     & 4.56$^{+0.89}_{-1.02}$  	 & 0 $\div$ 2$\pi$ 	& 4.83$^{+0.70}_{-0.80}$  	& 0 $\div$ 2$\pi$ \\
$\Delta m^2_{21}/10^{-5}\ [\mathrm{eV}^2]$ & 7.50$^{+0.19}_{-0.17}$ & 7.03 $\div$ 8.09 & 7.50$^{+0.19}_{-0.17}$ & 7.03 $\div$ 8.09\\
$\Delta m^2_{3\ell}/10^{-3}\ [\mathrm{eV}^2]$ & +2.524$^{+0.039}_{-0.040}$ & +2.407 $\div$ +2.643 & -2.514$^{+0.038}_{-0.041}$ & -2.635 $\div$ -2.399 \\
\bottomrule
\bottomrule
\end{tabular}
\caption{\small \it Central values $\pm$ the $1\sigma$ errors  and $3\sigma$ ranges for the neutrino mixing parameters as obtained in Ref. \cite{Esteban:2016qun} (available at the website \href{http://www.nu-fit.org}{\texttt{http://www.nu-fit.org}}). Note that in the last line $\ell =1$ for NO and $\ell = 2$ for IO. The analysis prefers a global minimum for NO with respect to the local minimum of IO,  $\Delta\chisquare = \chisquare_{\mathrm{IO}} - \chisquare_{\mathrm{NO}} = 0.83$.}
\label{tab:parameter_table}
\end{center}
\end{table}

The Planck Collaboration provides several likelihoods based on different assumptions among which we decide to use the following ones:
\begin{itemize}
 \item a conservative estimate ({\it{set-1}}) based on the set of data given by \PLANCKCON, which has $\sum_j m_j \leq 0.68$ eV at $95\% \ \mathrm{CL}$;
 \item a more aggressive one ({\it{set-2}}) based on \PLANCKAGG, which has $\sum_j m_j \leq 0.23$ eV at $95\% \ \mathrm{CL}$, 
 with  a maximum of the likelihood for $\sum_j m_j \sim 0.05\ \mathrm{eV}$.
\end{itemize}
The acronyms used above refer to the data on the temperature power spectrum (PLANCK TT), to the Planck polarization data in the low-$\ell$ temperature (lowP), 
to the data on Cosmic Microwave Background lensing reconstruction (Lensing); with Ext the constraints from Baryon Acoustic 
Oscillations, Joint Light-curve Analysis of supernovae and the Hubble constant are indicated.
For comparison purposes, we show in Tab. \ref{tab:planck_bound} the upper limits at 95\% CL on the sum of the neutrino masses for different datasets 
which also include the data from the temperature-polarization cross spectrum (TE) and those from the polarization power spectrum (EE).
\begin{table}[h!]
\begin{center}
\begin{tabular}{c  c  c  c c c c}
\toprule
\toprule
\bf  						& Dataset 	& + LowP &  + Lensing 	&  + Ext \\ 
\midrule	
\multirow{2}{*}{$\sum_j m_j\ [\mathrm{eV}]$}	& TT 		& 0.715  & 0.675 	& 0.234\\
						&TT, TE, EE 	& 0.492	 & 0.589 	& 0.194\\
\bottomrule
\bottomrule
\end{tabular}
\caption{\small \it Upper bound at 95\% confidence level on the sum of the neutrino masses (in eV) using the data of Ref. \cite{Ade:2015xua}.}
\label{tab:planck_bound}
\end{center}
\end{table}

With these likelihoods at our disposal, we employed the following procedure to accept or reject a given extraction of the set
of observables $p$ and Majorana phases (notice that 
 $\delta$ is not relevant because the Majorana phase $\beta$ hides any information on the Dirac $CP$ phase):
\begin{itemize}
 \item we first extract $\sin^2\theta_{12}$, $\sin^2\theta_{13}$, $\Delta m^2_{21}$ and $\Delta m^2_{3\ell}$ according to \eqref{like_from_deltachi};
 the Majorana phases $\alpha$ and $\beta$ are extracted according to a flat distribution in the interval $[0, 2\pi]$; 
 \item we then extract the value of $M=\sum_j m_j$ using the Planck data obtained from Fig. 30 in Ref. \cite{Ade:2015xua}; 
 for NO, if $M \leq \sqrt{\Delta m^2_{21}} + \sqrt{\Delta m^2_{31}}$ (or 
 $M \leq \sqrt{-(\Delta m^2_{21} + \Delta m^2_{32})} + \sqrt{-\Delta m^2_{32}}$ for IO), we reject  such an $M$  
 and extract a new value for the sum of the neutrino masses;
 \item once the value of $\sum_j m_j$ is accepted,  we compute the lightest neutrino mass $m_{\min}$ using the relations
 \begin{itemize}
  \item[$\circ$] $m_{\min} + \sqrt{m_{\min}^2 + \Delta m^2_{21}} + \sqrt{m_{\min}^2 + \Delta m^2_{31}} = \sum_j m_j$ for NO
  \item[$\circ$] $\sqrt{m_{\min}^2 - \Delta m^2_{21}-\Delta m^2_{32}} + \sqrt{m_{\min}^2 - \Delta m^2_{32}} + m_{\min}= \sum_j m_j$ for IO \,.
 \end{itemize}
 Notice that, unless the Planck distributions on $M$ are peaked around 0.06 eV assuming NO and 0.1 eV for IO (which is in fact not the case), this procedure penalizes very small values of $m_{\min}$.
\end{itemize}
Thus Eqs. \eqref{mbetabeta} and \eqref{mbeta_definition} are used to get the numerical values of  $m_{\beta}$ and $m_{\beta \beta}$.
We generate $\ord(10^6)$ realizations that satisfy the constraints discussed above. 
This order of magnitude is necessary to guarantee a $5\sigma$ coverage for the input parameters, as discussed in Ref. \cite{Benato:2015via}.

\section{PDF analysis}
\label{sec:results}
The procedure outlined above produces two-dimensional histograms in the planes $(m_{\min}, \mbetabeta)$ 
and $(\mbeta, \mbetabeta)$. In order to compute from them the PDF and CRs, we used the Kernel Density Estimate (KDE) approach \cite{Cranmer:2000du}.
Suppose we have a $d$-dimensional vector $\mbf{x}$ of observables of which we want to know the PDF, $f(\mbf{x})$, and suppose also that we have $N$ 
different realizations of the same observables $\{\mbf{t}_{j}\}_{j =1}^N$ obtained according to the procedure described above;
thus $f$ is estimated from
\begin{align}
 \hat{f}(\mbf{x}) = \frac{1}{N \prod_{k = 1}^d h_k}\sum_{j=1}^N \left[ \prod_{k = 1}^d \Ke\left(\frac{x^k - t_{j}^k}{h_k}\right)\right]\,,
\end{align}
where $h_k$ is the bandwidth of the $k$-th component of the vector $\mbf{x}$, whose estimate according to the Scott's rule of thumb \cite{Scott:1992} 
is given by
\begin{align}
\label{scott_rule}
 \hat{h}_k = \left(\frac{4}{d+4}\right)^{1/(d+4)}N^{-1/(d+4)} \sigma_k\,,
\end{align}
$\sigma_k$ being the standard deviation of the $k$-th observable $x^k$. The Scott's rule reduces the asymptotic expected value of the integrated square errors between the actual distribution $f$ 
and the estimated $\hat{f}$.\\
The positive function $\Ke$ is called {\it kernel} and 
must satisfy the normalization condition
\begin{align}
 \int_{\mathbb{R}^d} \de^dx\ \Ke(\mbf{x}) = 1 \qquad \Ke(\mbf{x}) \geq 0.
\end{align}
A simple but equally suited kernel is the Gaussian kernel, defined as:
\begin{align}
 \Ke(\mbf{x}) = \frac{1}{(2\pi)^{d/2}}\exp\left(-\frac{1}{2}|\mbf{x}|^2\right)\,,
 \end{align} 
that we estimate using the same algorithm of Ref. \cite{Crossfield:2010yq} \footnote{The original code is available at \href{https://people.ucsc.edu/~ianc/python/kdestats.html }{\texttt{https://people.ucsc.edu/\~{}ianc/python/kdestats.html }}. }, based on the modified \texttt{SciPy} function \texttt{gaussian\_kde} described in Ref. \cite{Fowlie:2016hew}.\\

\begin{figure}[h!]
\centering

 \includegraphics[scale=.5]{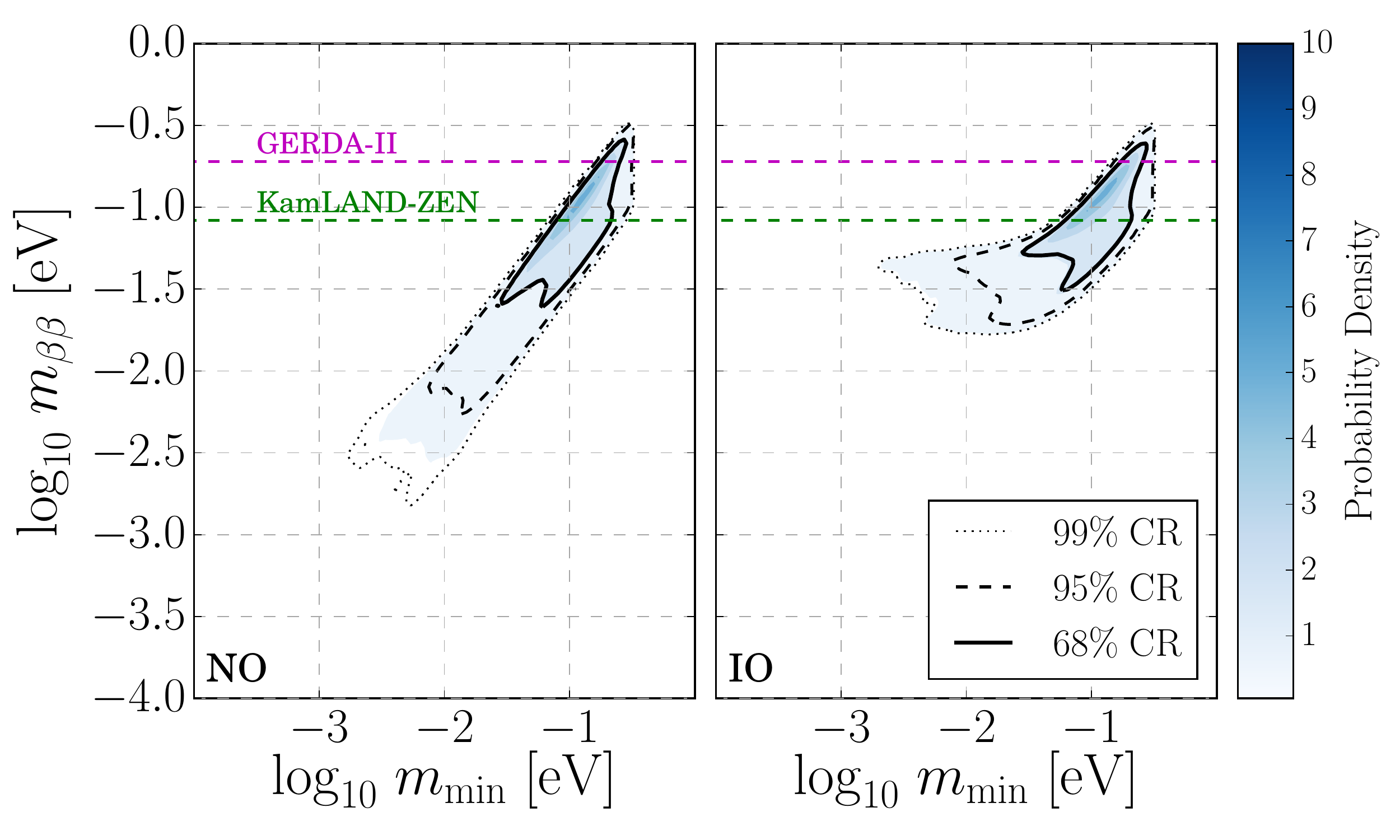}
  \caption{\small \it Reconstructed probability density in the $(m_{\min}, \mbetabeta)$ plane assuming NO (left panel) and IO (right panel) 
  using the {\it set-1} prior on $\sum_j m_j$. The credible regions are at 68\% (solid lines), 95\% (dashed lines) and 99\% (dotted lines) for 2 dof. The horizontal pink dashed line indicates the excluded region at 90\% CL assuming the $^{76}$Ge results \cite{Agostini:2017iyd}, while the green dashed line refers to 
  the $^{136}$Xe results \cite{KamLAND-Zen:2016pfg}.}
 \label{fig:mbb_mmin_LEM_planck_tt_lowP_lensing}
\end{figure}

\begin{figure}[h!]
\centering

 \includegraphics[scale=.5]{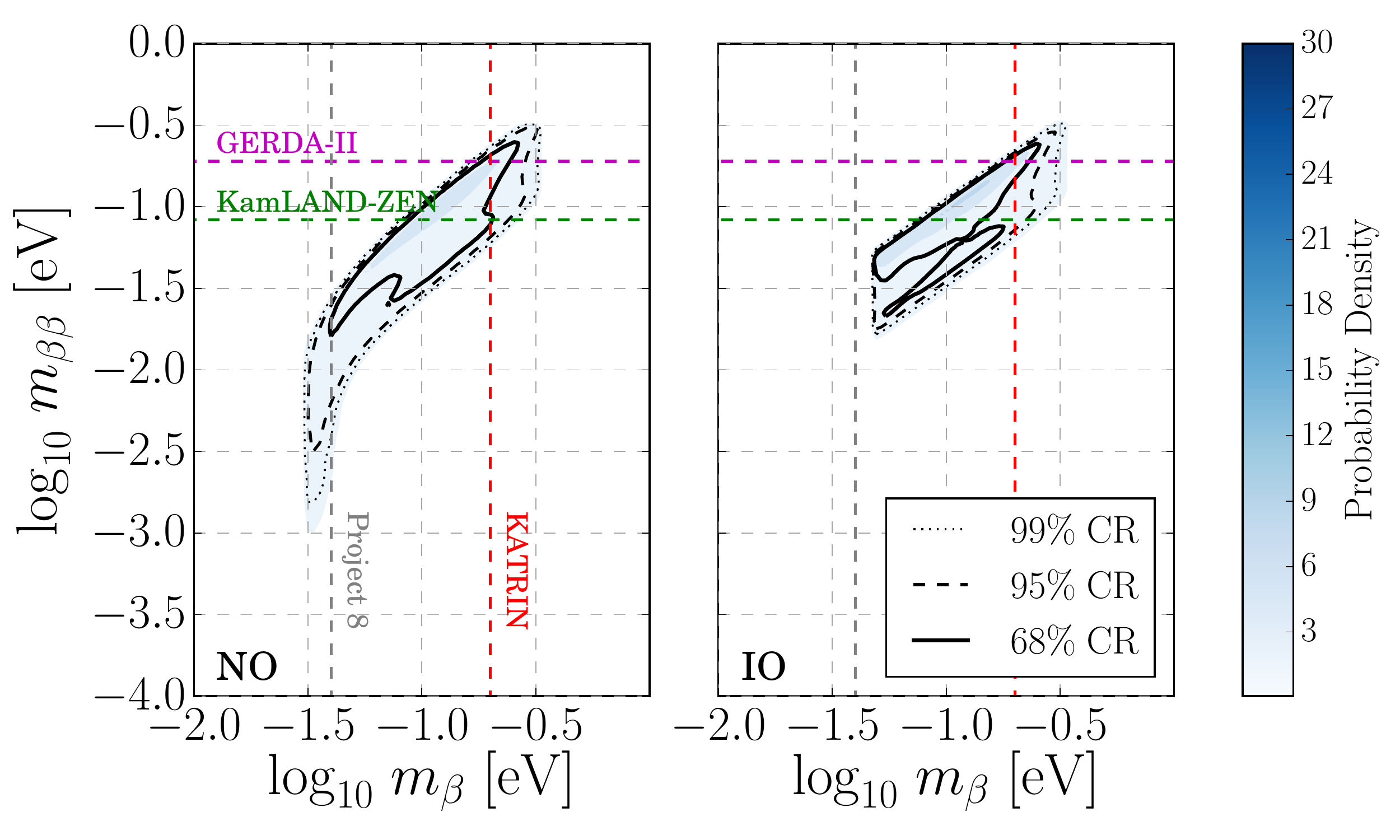}
  \caption{\small \it The same as Fig. \ref{fig:mbb_mmin_LEM_planck_tt_lowP_lensing} but in the $(\mbeta, \mbetabeta)$ plane. 
  With vertical red dashed lines we indicate the expected sensitivity of KATRIN \cite{Osipowicz:2001sq} and with vertical grey dashed lines the one of Project 8 \cite{Doe:2013jfe}.}
 \label{fig:mbb_mb_LEM_planck_tt_lowP_lensing}
\end{figure}

\begin{figure}[h!]
\centering

 \includegraphics[scale=.5]{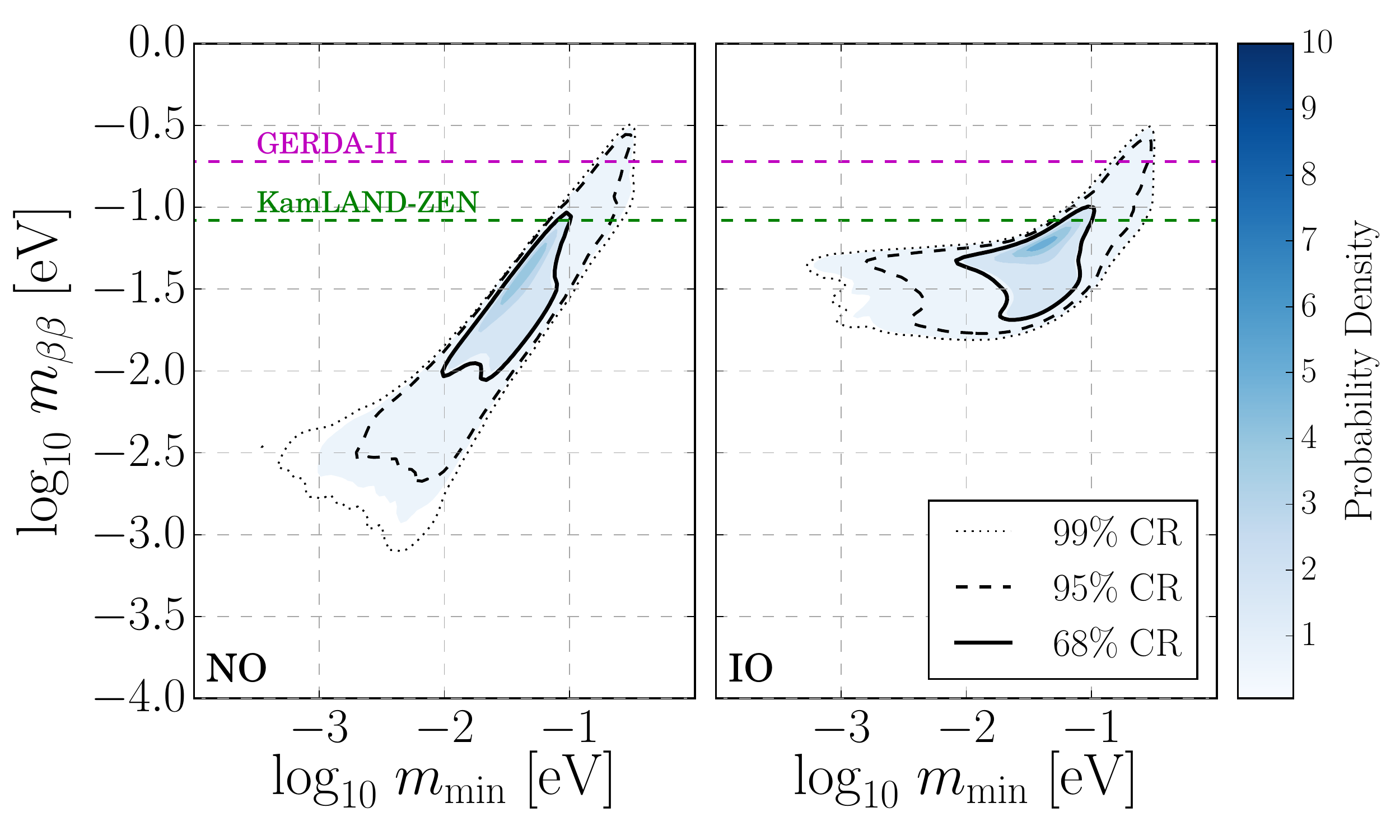}
  \caption{\small \it The same as Fig. \ref{fig:mbb_mmin_LEM_planck_tt_lowP_lensing}, but for {\it set-2}.}
 \label{fig:mbb_mmin_planck_tt_te_lowP_ext}
\end{figure}

\begin{figure}[h!]
\centering

 \includegraphics[scale=.5]{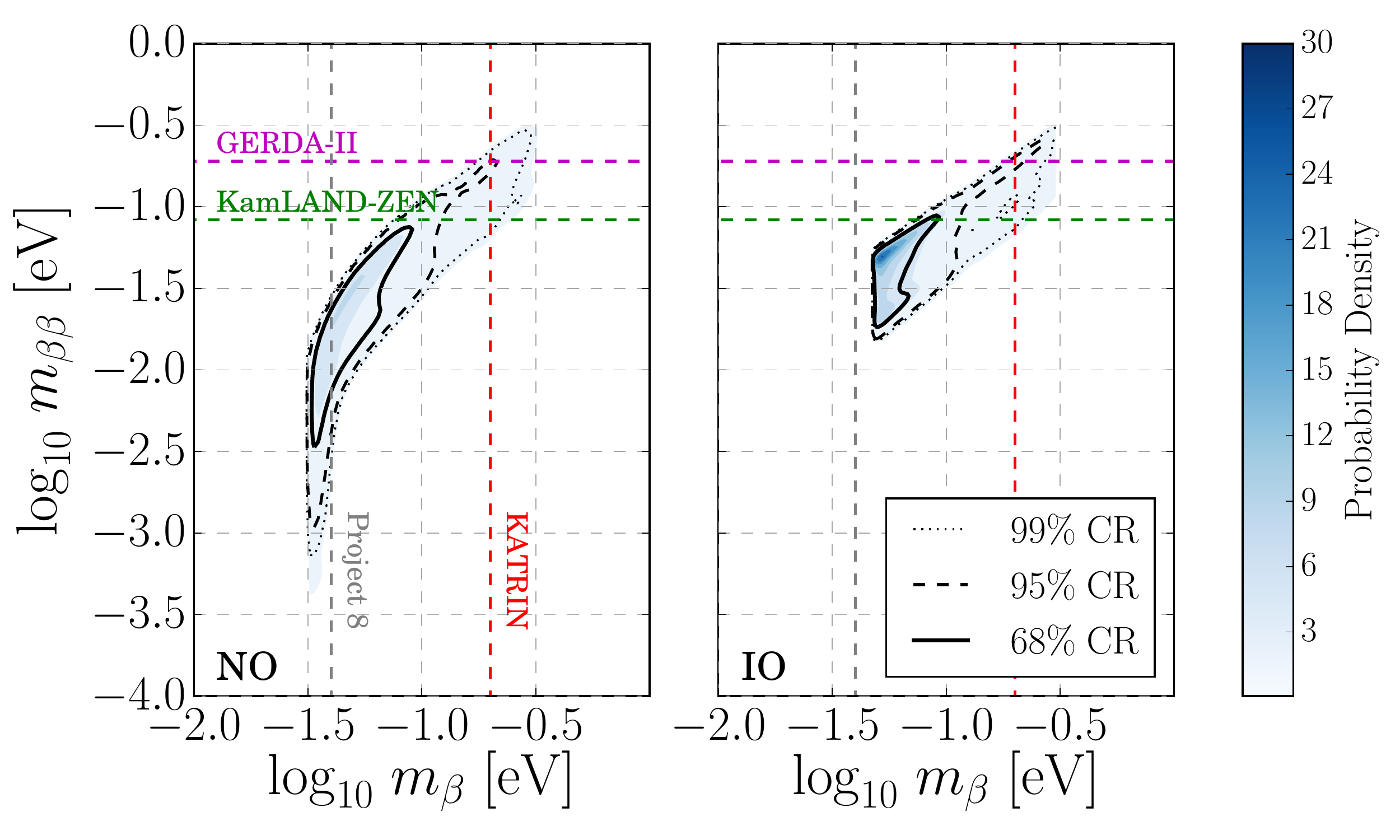}
  \caption{\small \it The same as Fig. \ref{fig:mbb_mb_LEM_planck_tt_lowP_lensing}, but for {\it set-2}.}
 \label{fig:mbb_mb_planck_tt_te_lowP_ext}
\end{figure}

The results for the PDFs as a function of $\log_{10} m_{\min}\ (\mbeta)$ and $\log_{10} \mbetabeta$ at the 68\%, 95\% and 99\% CRs obtained with the analysis performed using {\it set-1} for the sum of the neutrino masses are shown in Fig. \ref{fig:mbb_mmin_LEM_planck_tt_lowP_lensing} in the $(m_{\min}, \mbetabeta)$ plane and in Fig. \ref{fig:mbb_mb_LEM_planck_tt_lowP_lensing} in the 
$( m_{\beta},  \mbetabeta)$ plane. The analogous results for {\it set-2} are shown in Figs. \ref{fig:mbb_mmin_planck_tt_te_lowP_ext} and \ref{fig:mbb_mb_planck_tt_te_lowP_ext}. In all planes, the excluded region for $\mbetabeta$ is the area above the horizontal magenta dashed line, 
around $\mbetabeta \geq 0.19\ \mathrm{eV}$ \cite{Dell'Oro_nufact} obtained using the 90\% CL limit on the half-life of $^{76}$Ge, $T^{0\nu}_{1/2}(^{76}\mathrm{Ge}) > 5.2 \times 10^{25}\ \mathrm{years}$, in the preliminary analysis of GERDA phase II \cite{Agostini:2017iyd}. A recent result using the $^{136}$Xe, $T^{0\nu}_{1/2}(^{136}\mathrm{Xe}) > 1.07 \times 10^{26}\ \mathrm{years}$ at $90\% \ \mathrm{CL}$ obtained by the KamLAND-ZEN experiment \cite{KamLAND-Zen:2016pfg}, gives the lower bound indicated with green dashed lines, which excludes the region $\mbetabeta \geq 0.083\ \mathrm{eV}$ \cite{Dell'Oro_nufact}. The bounds we quote for $\mbetabeta$ are obtained according to Ref. \cite{Dell'Oro:2016dbc}, where the Authors used the results of Ref. \cite{Kotila:2012zza} for the phase-space factor and those of Ref. \cite{Barea:2015kwa} for the nuclear matrix elements. In our analysis we fixed  the axial coupling constant of the nucleon $g_A=1.269$. 
 We also outline that the large uncertainties associated to the nuclear matrix elements $ \mathcal{M}(A, Z)$ can modify the prediction  for decay amplitude $ \mathcal{A}(0\nu\beta\beta)$; however, the impact of such effects are beyond the scope of this paper and will not be analyzed in the following.
For the other observable, $\mbeta$, the red vertical dashed line indicates the expected sensitivity of the KATRIN experiment 
($0.2\ \mathrm{eV}$ at $90\% \ \mathrm{CL}$ \cite{Osipowicz:2001sq}, see \href{https://www.katrin.kit.edu/128.php}{\texttt{https://www.katrin.kit.edu/128.php}}) and the grey vertical dashed line  the expected sensitivity of the Project 8 experiment 
($4 \times 10^{-2}\ \mathrm{eV}$ at $90\% \ \mathrm{CL}$ \cite{Doe:2013jfe}) which has been especially designed to probe the whole IO parameter space.
Notice that the most stringent upper limit on $m_\beta$ has been obtained by the Mainz and Troitzk experiments, $\mbeta \leq 2.05\ \mathrm{eV}$ at $95\% \ \mathrm{CL}$ \cite{Lobashev:2003kt, Aseev:2011dq}. 
To better compare the different cosmological datasets we use the same scale for the PDF densities (which are normalized to one). In the $(m_{\min},  \mbetabeta)$ plane the maximum is fixed to be 10, while in the $( \mbeta,  \mbetabeta)$ plane it is 30.

\noindent
A close inspection at Fig. \ref{fig:mbb_mmin_LEM_planck_tt_lowP_lensing} shows that a large portion of the  68\% CR for $\mbetabeta$, the one corresponding to large $m_{\min}$, is already excluded by the KamLAND-ZEN data, for both hierarchies. In practice, this is the consequence of having  a non-negligible probability that $\sum_j m_j \gtrsim 0.5\ \mathrm{eV}$, therefore the value of $m_{\min}$ can be sufficiently large to approach ${\cal O}(0.1)$. On the other hand, {\it set-2} 
relaxes this constraint and the probability density is centered around smaller $m_{\min}$ (and consequently smaller $\mbetabeta$).
For both cases, the cosmological bounds on the sum of neutrino masses implies that for NO and IO the low mass region for $\mbetabeta$ is strongly disfavoured. 

The interesting features of Figs. \ref{fig:mbb_mb_LEM_planck_tt_lowP_lensing} and  \ref{fig:mbb_mb_planck_tt_te_lowP_ext} is that, for both assumptions on 
the sum of the neutrino masses, the Project 8 experiment would be able to probe almost the whole allowed regions for $m_\beta$ at 99\% level whereas KATRIN shows only a modest ability to probe the largest possible values of $m_\beta$, around ${\cal O}(10^{-2}-10^{-1})$ eV.

Instead of discussing the effects of the cosmological bounds on the effective masses, one can also adopt an opposite point of view, asking what would be the effect on $\sum_j m_j$ of a possible measure of $\mbetabeta$ at the  new generation of experiments, see Refs. \cite{Giuliani:2012zu,Bilenky:2014uka,Dell'Oro:2016dbc}.
As an example, we can explore the situation that a signal for the $0\nu\beta\beta$-decay is observed at the (near future) CUORE or (next-to-near future) nEXO experiments. 
Following the discussion of Ref. \cite{Dell'Oro:2016dbc} we assume an optimistic scenario where a signal is in the expected 90\% experimental sensitivity region, that is $\mbetabeta =  0.073 \pm 0.008\ \mathrm{eV}$ (assuming $g_A = 1.269$) for CUORE \cite{Artusa:2014lgv}. 
Similar values can also be achieved by GERDA Phase-II \cite{Brugnera:2013xma}, MAJORANA-D \cite{Abgrall:2013rze} and NEXT \cite{Gomez-Cadenas:2013lta} experiments, so our discussion applies equally well to a large number of possible future experiments.
In the case of nEXO experiment \cite{Pocar:2015mrz}, we set $\mbetabeta =  0.011 \pm 0.001\ \mathrm{eV}$, which is below the IO region.

\begin{figure}[h!]
\centering

 \includegraphics[scale=.5]{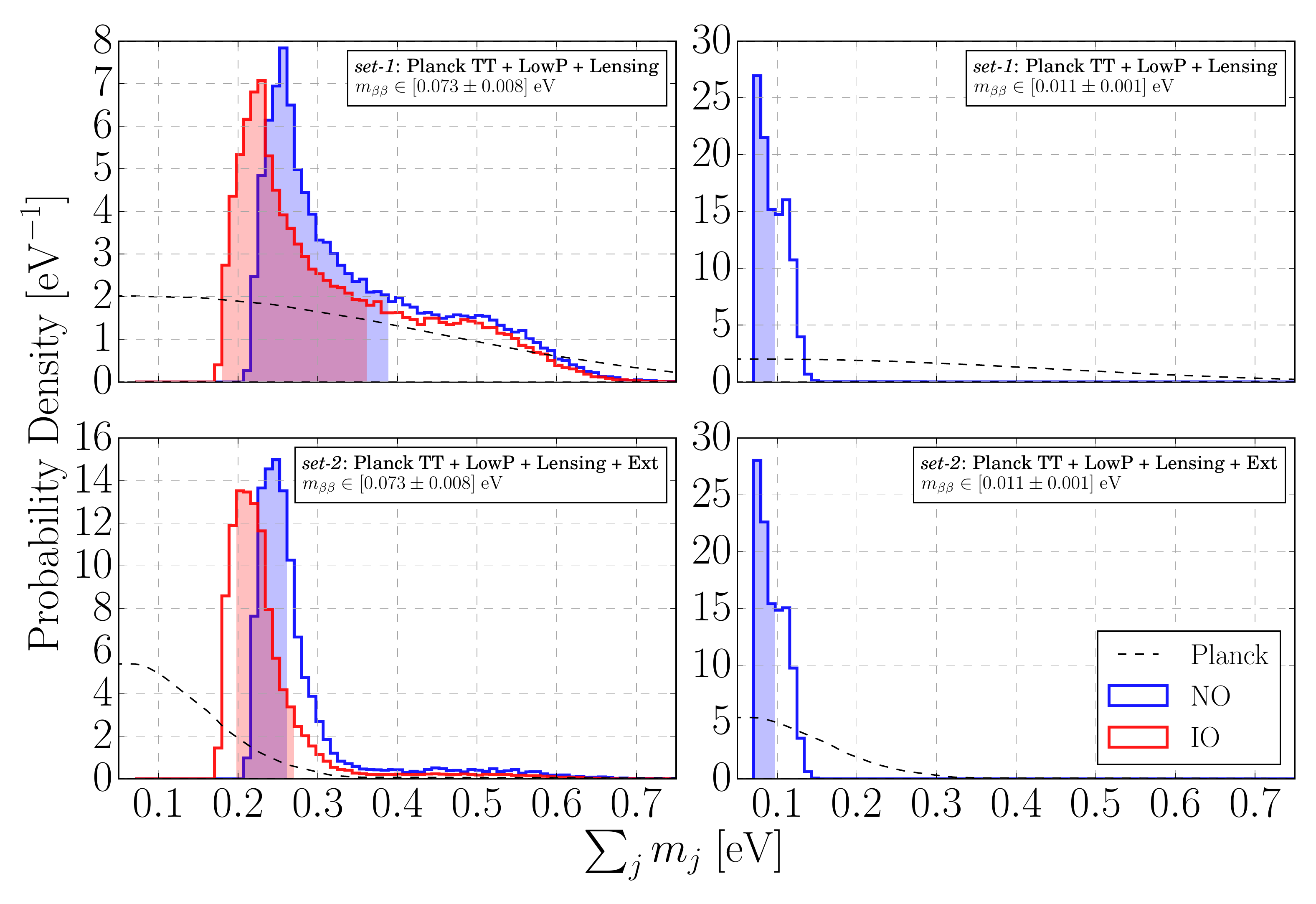}
  \caption{\small \it Frequency of $\sum_j m_j$ for an assumed $\mbetabeta =  0.073 \pm 0.008\ \mathrm{eV}$ (left panels) and  $\mbetabeta =  0.011 \pm 0.001\ \mathrm{eV}$ (right panels), for NO (blue) and IO (red). The black dashed lines are the Planck PDFs: {\it set-1} in the upper panels and {\it set-2} in the lower panels. The darkest areas under the histograms are the 68\% credible regions obtained from the cumulant distributions.}
 \label{fig:1d_dimension_november2016}
\end{figure}

The results of our finding are shown in Fig. \ref{fig:1d_dimension_november2016} where the frequency of the sum of light neutrino masses (histograms normalized to 1), after the constraints coming from $\mbetabeta$, is displayed. With black dashed lines we also show the Planck PDFs for {\it set-1} (upper panels) and {\it set-2} (lower panels).
%
In the first  column of the plot, which refers to the case $\mbetabeta =  0.073 \pm 0.008\ \mathrm{eV}$,
we clearly see that there exists a cutoff in the distribution in the low mass region due to the fact that $\mbetabeta$ cannot be arbitrarily small, with maxima 
around $\sum_j m_j\sim {\cal O}(0.2-0.3)$ eV for both {\it set-1} and {\it set-2} and for both hierarchies (NO in blue and  IO in red).
On the other hand, in the high mass region the distributions essentially follow the shape of the Planck priors since the assumed values of $\mbetabeta$ do not impose 
strong constraints on $m_{\min}$. 

If we assume  a positive signal at the nEXO experiment $\mbetabeta =  0.011 \pm 0.001\ \mathrm{eV}$, second column of Fig. \ref{fig:1d_dimension_november2016},
we see that we cannot distinguish among different Planck datasets since the bound on the $0\nu\beta\beta$-decay effective mass constraints 
$\sum_j m_j$ to be of $\ord(0.1)\ \mathrm{eV}$.

\section{Discussion and conclusions}
\label{discussion}
\begin{figure}[h!]
\centering
 \includegraphics[scale=.5]{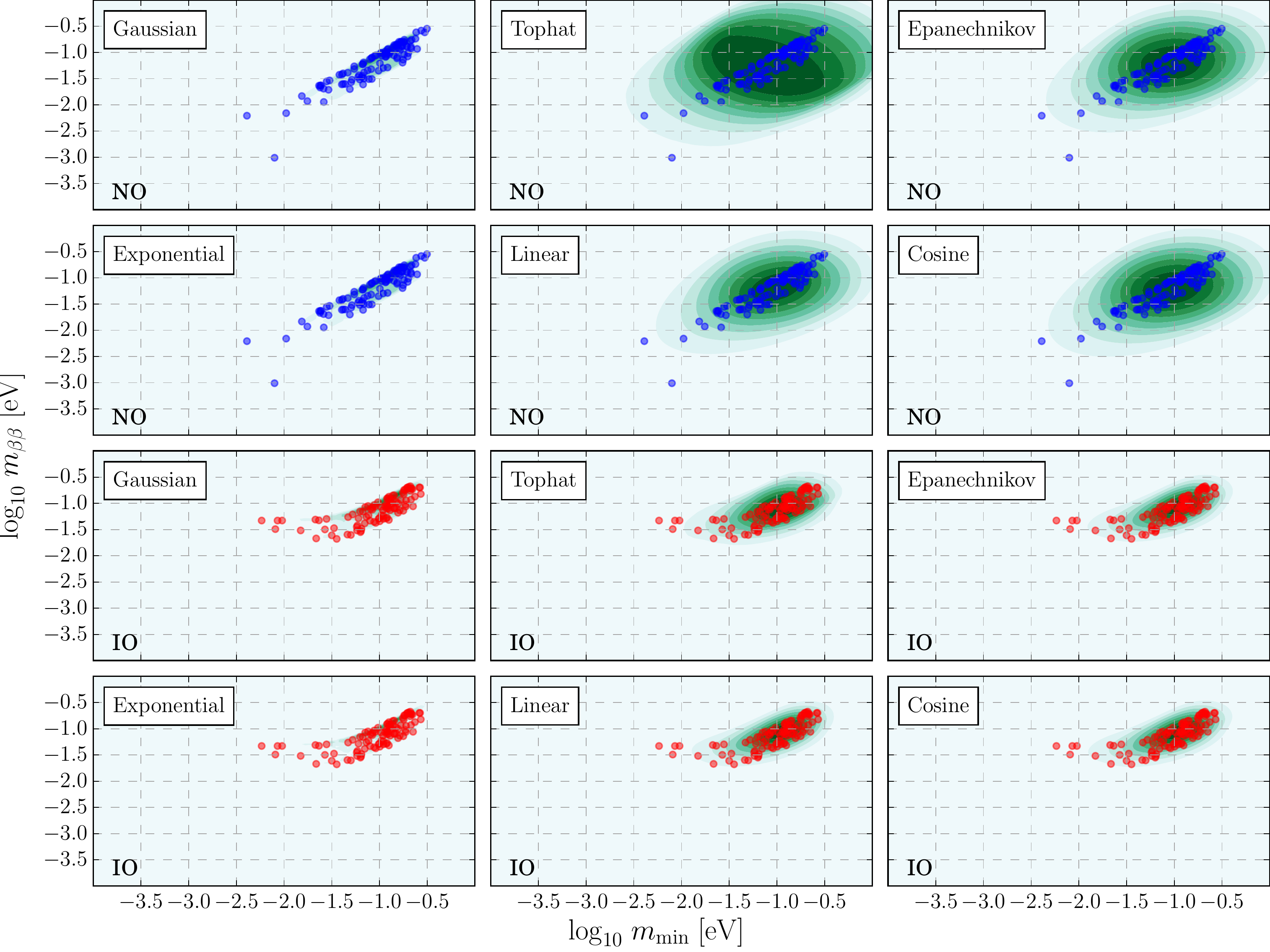}
  \caption{\small \it  PDFs in the plane $(m_{\min},\mbetabeta)$ obtained from the KDE analysis (green shaded areas) for NO and IO datasets; the blue (red) 
  points are a sample of data obtained in the numerical scan.}
 \label{fig:compare_kernels}
\end{figure}

At first sight, the results described above seem to depend on the choice of the kernel used to estimate the PDFs. 
However, we have checked that adopting  different functions $\Ke$ the CL regions are not altered in a significant manner. We test different kernels, provided by the \texttt{scikit-learn} package \cite{scikit-learn}: 
\begin{itemize}
\item {\it Gaussian} $\Ke(x; h) \propto \exp(-x^2/2h)$\,;
 \item {\it tophat} $\Ke(x; h) \propto 1$ for $|x| \leq h$\,;
 \item {\it Epanechnikov} $\Ke(x; h) \propto 1- x^2/h^2$\,;
 \item {\it exponential} $\Ke(x; h) \propto \exp(-|x|/h)$\,;
 \item {\it linear} $\Ke(x; h) \propto 1-x/h$ for $|x| \leq h$\,;
 \item {\it cosine} $ \Ke(x; h) \propto \cos(\pi x/2 h)$\,.
\end{itemize}
The check is performed adopting the {\it $k$-fold cross-validation} approach, proposed in Refs. \cite{Mosteller:68, Stone:1974};
in few words, the sample of extracted points is split into $k$ smaller sets; of them, $k-1$ sets are used to estimate $f(\mbf{x})$ according to a given 
{\it kernel} and the  resulting model is then validated on the remaining part of the dataset.
In Tab. \ref{tab:kernel_accuracy} we show our result for the cross-validation analysis with $m_{\rm min}$ and $\mbetabeta$ as independent variables (similar results can be achieved for $\mbeta$ and $\mbetabeta$): we analyze ten subsets with $N = \{1000, 5000, 10000, 50000\}$ points, then we average the results. In order to investigate possible overfitting effects, each subset has been divided into two parts: a \emph{train} ($N_{\rm train} \approx 0.6 N$) and a \emph{test} ($N_{\rm test} = N - N_{\rm train}$) set. We estimate the best bandwidth $\hat{h}$ using twenty $k$-folds in the train dataset. The error ${\cal E_{\rm set}}$ between the actual distribution and the kernel estimate is defined as:
\begin{align}
\label{error_cv}
{\cal E}_{\rm set} = \sqrt{ \frac{1}{N_{\rm set}}  \sum_j^{N_{\rm set}^{1/2}}\sum_k^{N_{\rm set}^{1/2}}\left[f(\mbf{x}_{j, k}) - \hat{f}(\mbf{x}_{j, k})\right]^2 } \,,
\end{align}
where {\it set} can be train or test-set. The actual distribution can be obtained from the two dimensional density histogram. We assume for the histogram $N_{\rm set}^{1/2} \times N_{\rm set}^{1/2}$ bins. Notice that the normalization factor $N^{-1/2}_{\rm set}$ in the error \eqref{error_cv} is necessary to compare datasets with different dimensions.\\
In Fig. \ref{fig:compare_kernels} we show our results of $\hat{f}(\mbf{x})$ for the {\it{set-1}} prior on the sum of neutrino masses
and for all kernels introduced above (green shaded area). The PDFs are superimposed on a subset of $5 \times 10^3$ points. 
As we can see, the Gaussian kernel 
as well as the exponential one correctly reproduce the testing dataset for both orderings (for these two cases, the green areas are concentrated 
below the points and they do not appear in the graphs). 
For the other kernels, the agreement does not appear to be as good as for the previous ones, since  the PDFs extend over regions outside 
the subset of points. In particular,  in Tab. \ref{tab:kernel_accuracy}  we observe that the errors ${\cal E}_{{\rm set}}$ of the Gaussian and the exponential kernels are
roughly  one half those of  the other kernels.

The cross-validation procedure is also useful to compute the best bandwidth $\hat{h}$ that minimizes the residual error between the predictions and the actual values of the sample points. Our findings are compatibles with the Scott's rule defined in \eqref{scott_rule}, see Tab. \ref{tab:scott_rule_summary} for a summary of the bandwidths computed using the same data of the cross-validation analysis. Notice that in the cross-validation a single bandwidth is estimated for each kernel.
For the {\it{set-2}} our conclusions remain unaltered: the Gaussian  
and the exponential kernels reproduce the training dataset with a good accuracy.

\begin{table}[h!]
\begin{center}
\resizebox{\textwidth}{!}{  
\begin{tabular}{c c  c c c  c c c   c c c  c c c }
\toprule
\toprule
& & \multicolumn{6}{c}{{\it set-1} } &\multicolumn{6}{c}{{\it set-2} }\\
& & \multicolumn{3}{c}{Normal Ordering} & \multicolumn{3}{c}{Inverted Ordering} &\multicolumn{3}{c}{Normal Ordering} & \multicolumn{3}{c}{Inverted Ordering}\\
\midrule
$N$ & Kernel & $\hat{h}$ & ${\cal E}_{{\rm train}}$ & ${\cal E}_{{\rm test}}$ & $\hat{h}$ & ${\cal E}_{{\rm train}}$ & ${\cal E}_{{\rm test}}$ & $\hat{h}$ & ${\cal E}_{{\rm train}}$ & ${\cal E}_{{\rm test}}$ & $\hat{h}$ & ${\cal E}_{{\rm train}}$ & ${\cal E}_{{\rm test}}$\\ 
\midrule
\multirow{6}{*}{\rotatebox[origin=c]{90}{\parbox[t]{1cm}{\centering 1000}}} & Gaussian &  0.060 & 0.204 & 0.225 & 0.068 & 0.378 & 0.421 & 0.076 & 0.146 & 0.181 & 0.058 & 0.350 & 0.430\\
& Tophat & 0.537 & 0.403 & 0.397 & 0.829 & 0.606 & 0.630 & 0.759 & 0.335 & 0.346 & 0.423 & 0.531 & 0.594 \\
& Epanechnikov & 0.537 & 0.367 & 0.360 & 0.829 & 0.568 & 0.593 & 0.759 & 0.305 & 0.316 & 0.423 & 0.491 & 0.556\\
& Exponential & 0.026 & 0.189 & 0.233 & 0.026 & 0.363 & 0.415 & 0.032 & 0.142 & 0.178 & 0.023 & 0.328 & 0.423\\
& Linear &  0.537 & 0.354 & 0.347 &  0.829 & 0.553 & 0.580 &  0.759 & 0.293 & 0.306 &  0.423 & 0.477 & 0.543\\
& Cosine &  0.537 & 0.465 & 0.459 & 0.829 & 0.695 & 0.717 & 0.759 & 0.363 & 0.375 & 0.423 & 0.625 & 0.692\\
\midrule
\multirow{6}{*}{\rotatebox[origin=c]{90}{\parbox[t]{1cm}{\centering 5000}}} & Gaussian & 0.043 & 0.145 & 0.167 & 0.033 & 0.301 & 0.326 & 0.046 & 0.121 & 0.144 & 0.042 & 0.299 & 0.304\\
& Tophat & 0.778 & 0.365 & 0.386 &  0.483 & 0.582 & 0.577 & 0.759 & 0.310 & 0.343 & 0.803 & 0.515 & 0.503\\
& Epanechnikov & 0.778 & 0.338 & 0.359 & 0.483 & 0.544 & 0.539 & 0.759 & 0.289 & 0.320 & 0.803 & 0.487 & 0.476\\
& Exponential & 0.016 & 0.125 & 0.153 &  0.014 & 0.281 & 0.323& 0.020 & 0.101 & 0.127 & 0.016 & 0.256 & 0.277\\
& Linear & 0.778 & 0.329 & 0.349 & 0.483 & 0.523 & 0.525 & 0.759 & 0.280 & 0.311 & 0.803 & 0.476 & 0.466 \\
& Cosine & 0.778 & 0.394 & 0.417 & 0.483 & 0.654 & 0.653  & 0.759 & 0.333 & 0.367 & 0.803 & 0.588 & 0.577 \\
\midrule
\multirow{6}{*}{\rotatebox[origin=c]{90}{\parbox[t]{1cm}{\centering 10000}}} & Gaussian & 0.031 & 0.137 & 0.146 & 0.025 & 0.295 & 0.306 &  0.035 & 0.105 & 0.127 & 0.029 & 0.272 & 0.284\\
& Tophat & 0.583 & 0.352 & 0.356 &  0.356 & 0.567 & 0.554 & 0.692 & 0.289 & 0.319 &  0.607 & 0.517 & 0.523\\
& Epanechnikov &  0.583 & 0.326 & 0.331 & 0.356 & 0.531 & 0.518& 0.692 & 0.270 & 0.299 & 0.607 & 0.489 & 0.497\\
& Exponential & 0.013 & 0.111 & 0.130 & 0.012 & 0.272 & 0.299 & 0.015 & 0.084 & 0.113 & 0.012 & 0.232 & 0.255\\
& Linear & 0.583 & 0.317 & 0.322 &  0.356 & 0.519 & 0.505 & 0.692 & 0.262 & 0.290 &  0.607 & 0.479 & 0.486\\
& Cosine & 0.583 & 0.391 & 0.399 &  0.356 & 0.652 & 0.642 & 0.692 & 0.318 & 0.350 & 0.607 & 0.578 & 0.586\\
\midrule
\multirow{6}{*}{\rotatebox[origin=c]{90}{\parbox[t]{1cm}{\centering 50000}}} & Gaussian & 0.019 & 0.118 & 0.115 & 0.018 & 0.252 & 0.274 & 0.019 & 0.083 & 0.089 & 0.019 & 0.238 & 0.252\\
& Tophat & 0.544 & 0.342 & 0.340 &  0.865 & 0.584 & 0.607 & 0.495 & 0.267 & 0.275 &  0.809 & 0.516 & 0.536 \\
& Epanechnikov &  0.544 & 0.320 & 0.316 &   0.865 & 0.557 & 0.579& 0.692 & 0.270 & 0.299 & 0.607 & 0.489 & 0.497\\
& Exponential & 0.008 & 0.089 & 0.097 &  0.007 & 0.222 & 0.251 & 0.009 & 0.068 & 0.079 & 0.007 & 0.190 & 0.212 \\
& Linear & 0.544 & 0.312 & 0.308 & 0.865 & 0.546 & 0.568& 0.495 & 0.241 & 0.248 &  0.809 & 0.484 & 0.502 \\
& Cosine & 0.544 & 0.369 & 0.367 & 0.865 & 0.630 & 0.655 & 0.495 & 0.297 & 0.306 & 0.809 & 0.562 & 0.582\\
\bottomrule
\bottomrule
\end{tabular}
}
\caption{\small \it Mean estimated bandwidth and mean errors for the train and test datasets assuming NO or IO. The results are obtained using a sample of $N$ points performing twenty $k$-folds cross-validation for the train subset. The error is defined in \eqref{error_cv}.}
\label{tab:kernel_accuracy}
\end{center}
\end{table}

\begin{table}[h!]
\begin{center}
\begin{tabular}{c   c c c c  c c c c}
\toprule
\toprule
& \multicolumn{4}{c}{{\it set-1}} & \multicolumn{4}{c}{{\it set-2}}\\
& \multicolumn{2}{c}{Normal Ordering} & \multicolumn{2}{c}{Inverted Ordering} & \multicolumn{2}{c}{Normal Ordering} & \multicolumn{2}{c}{Inverted Ordering}\\
\midrule
$N$ & $h_{m_{\rm min}}$ & $h_{m_{\beta \beta}}$ & $h_{m_{\rm min}}$ & $h_{m_{\beta \beta}}$ & $h_{m_{\rm min}}$ & $h_{m_{\beta \beta}}$ & $h_{m_{\rm min}}$ & $h_{m_{\beta \beta}}$\\ 
\midrule
1000  & 0.396 & 0.407 & 0.393 & 0.252 & 0.434 & 0.416 & 0.466 & 0.203 \\
5000  & 0.311 & 0.316 & 0.300 & 0.193 & 0.332 & 0.319 & 0.344 & 0.155 \\
10000 & 0.279 & 0.283 & 0.267 & 0.173 & 0.292 & 0.284 & 0.314 & 0.138 \\
50000 & 0.213 & 0.217 & 0.205 & 0.132 & 0.223 & 0.217 & 0.238 & 0.105 \\
\bottomrule
\bottomrule
\end{tabular}
\caption{\small \it Mean values of the bandwidths evaluated for the Gaussian kernel using the Scott's rule defined in \eqref{scott_rule} and the same data of Tab. \ref{tab:kernel_accuracy}.}
\label{tab:scott_rule_summary}
\end{center}
\end{table}

In conclusions, we have shown that the KDE method is an efficient tool to evaluate the PDFs of interesting physical observables. 
We have concentrated our efforts on two observables related to neutrino physics, namely the effective neutrino masses $m_{\beta\beta}$ and $m_\beta$ which will help to reveal the true nature of neutrinos and the values of their absolute masses. For them, we have computed the Credible Regions using the available PDFs on the mixing angles and mass differences, with the additional constraints coming from cosmological data on the sum of the neutrino masses. We found that the reconstructed probability densities strongly depend on the assumed set of cosmological data and, in particular, 
for  $\sum_j m_j \leq 0.23$ eV at $95\% \ \mathrm{CL}$  the bulk of the probability densities are below the current bounds on the analyzed observables. This conclusion remains qualitatively unaffected if one uses a different choice of the kernel function.

\section*{Acknowledgements}

We are indebted with Carlo Giunti for useful discussion about the neutrino effective masses and Andrew Fowlie for sharing his code to compute in a different way the 1D and 2D posterior PDFs. 



\bibliographystyle{elsarticle-num} 
\bibliography{manuscript}
\end{document}